\documentclass[aps,prl,twocolumn,showpacs,preprintnumbers,superscriptaddress,floatfix,amsmath,amssymb,amsfonts,10pt,longbibliography]{revtex4-1}

\usepackage{amsmath,amsfonts,amssymb}
\usepackage{graphicx}
\usepackage{color}

\usepackage[colorlinks=true,
            linkcolor=blue,
            urlcolor=blue,
            citecolor=blue]{hyperref}
\usepackage[sf]{subfigure}
\usepackage{MnSymbol}

\makeatletter
\@tfor\next:=abcdefghijklmnopqrstuvwxyzABCDEFGHIJKLMNOPQRSTUVWXYZ\do{%
  \def\command@factory#1{%
    \expandafter\def\csname bf#1\endcsname{\mathbf{#1}}
  }
 \expandafter\command@factory\next
}
\@tfor\next:=abcdefghijklmnopqrstuvwxyzABCDEFGHIJKLMNOPQRSTUVWXYZ\do{%
  \def\command@factory#1{%
    \expandafter\def\csname bb#1\endcsname{\mathbb{#1}}
  }
 \expandafter\command@factory\next
}
\@tfor\next:=abcdefghijklmnopqrstuvwxyzABCDEFGHIJKLMNOPQRSTUVWXYZ\do{%
  \def\command@factory#1{%
    \expandafter\def\csname cal#1\endcsname{\mathcal{#1}}
  }
 \expandafter\command@factory\next
}
\@tfor\next:=abcdefghijklmnopqrstuvwxyzABCDEFGHIJKLMNOPQRSTUVWXYZ\do{%
  \def\command@factory#1{%
    \expandafter\def\csname til#1\endcsname{\widetilde{#1}}
  }
 \expandafter\command@factory\next
}
\makeatother

\newcounter{ParagraphCounter}
\newcommand{\para}[1]{%
\noindent%
\fbox{%
\textsf{\textbf{%
\arabic{ParagraphCounter}}}}%
\stepcounter{ParagraphCounter}%
\;%
}

\newcommand{\up}[0]{\uparrow}
\newcommand{\dn}[0]{\downarrow}

\renewcommand{\para}[1]{}

\begin{document}

\title{Emergent topological superconductivity at nematic domain wall of FeSe}
\author{Kyungmin Lee}
\affiliation{Department of Physics, The Ohio State University, Columbus, OH 43210, USA}
\author{Eun-Ah Kim}
\affiliation{Department of Physics, Cornell University, Ithaca, New York 14853, USA}
\date{\today}

\begin{abstract}
One dimensional hybrid systems play an important role in the search for topological superconductivity.
Nevertheless, all one dimensional hybrid systems so far have been externally defined.
Here we show that one-dimensional domain wall in a nematic superconductor can serve as an emergent hybrid system in the presence of spin-orbit coupling.
As a concrete setting we study the domain wall between nematic domains in FeSe, which is well established to be a nematic superconductor.
We first show on the symmetry grounds that spin-triplet pairing can be induced at the domain wall by constructing a Ginzburg-Landau theory.
We then demonstrate using Bogoliubov-de Gennes approach that such nematic domain wall supports zero energy bound states which would satisfy Majorana condition. 
Well-known existence of these domain walls at relatively high temperatures, which can in principle be located and investigated with scanning tunneling microscopy, presents new opportunities for a search for realization of Majorana bound states. 
\end{abstract}
\maketitle

\para{}%-------------------------------------------------------------
The realization that one-dimensional hybrid systems with spin-orbit coupling (SOC)~\cite{mourik_signatures_2012,deng_majorana_2016}  or magnetism~\cite{nadj-perge_observation_2014} can offer a new arena for topological superconductivity [see Ref.~\onlinecite{0034-4885-75-7-076501} and references therein] has led to renewed interest in one-dimensional superconductors.
Moreover, broader appreciation of edge state properties of Dirac systems such as Weyl semimetal~\cite{murakami_phase_2007,wan_topological_2011}, graphene~\cite{fujita_peculiar_1996}, and nodal superconductors including a $d$-wave superconductor \cite{hu_midgap_1994,wang_topological_2012} have emerged.
Unfortunately, however, most of these systems require rather special conditions operating at extremely low temperatures.
On the other hand, little attention has been paid to the fact that a domain wall in a superconductor with additional $\mathbb{Z}_2$ symmetry breaking could form a new type of hybrid systems, although nematic order in superconducting phase is common~\cite{RevModPhys.87.457,fernandes_what_2014}.
In iron-based superconductors, in particular, robust signatures of nematic phase transition has been detected and imaged~\cite{chu_in_2010,chuang_nematic_2010}.
Boundaries between nematic domains, embedded in a spin-orbit coupled system, provide a new possibility towards realizing a novel one-dimensional superconductor.
Motivated by this, we consider a new emergent hybrid situation of nematic domain wall in FeSe.

\para{}%-------------------------------------------------------------
FeSe has generated much interest as a superconductor which exhibits nematicity without additional complication from magnetic order~\cite{wang_interface_2012,liu_electronic_2012}.
Through real space probes, boundaries between two such nematic domains have been observed~\cite{kalisky_stripes_2010,song_direct_2011,watashige_evidence_2015}.
In this letter, we study the structure of superconducting pairing at nematic domain walls using symmetry analysis, and argue that spin-triplet pairing can be induced through spin-orbit coupling.
Furthermore, we show, through Bogoliubov-de Gennes approach, that the ends of domain walls can support zero energy bound states.
We also remark on the connection between our results and the recent works on the edge states of nodal superconductor~\cite{potter_edge_2014,hofmann_edge_2016}.

\begin{table}
\begin{tabular}{c|ccc}
\toprule
symmetry        & tetragonal & orthorhombic & domain wall   \\
operation       & phase      & phase        & $[1\bar{1}0]$ \\
\hline
$2C_4$          & $\circ$    & $\times$     & $\times$      \\
$C_2$           & $\circ$    & $\circ$      & $\times$      \\
$2\sigma_v$     & $\circ$    & $\circ$      & $\times$      \\
$\sigma_d(x-y)$ & $\circ$    & $\times$     & $\circ$       \\
$\sigma_d(x+y)$ & $\circ$    & $\times$     & $\times$      \\
\hline
point group     & $C_{4v}$   & $C_{2v}$     & $C_s$         \\
\botrule
\end{tabular}
\caption{\label{tab:broken}%
Point group symmetries in different phases.
$\circ$ and $\times$ respectively denote good and broken symmetry operations.
The linear polynomial $p$ in $\sigma_d(p)$ defines the mirror plane by $p = 0$.
}
\end{table}

\para{}%-------------------------------------------------------------
The superconducting gap structure of FeSe is still an unsettled issue.
The two most widely discussed pairing symmetries of iron-based superconductors are $s$-wave and $d_{x^2-y^2}$-wave~\cite{stewart_superconductivity_2011}.
In the tetragonal phase of FeSe, the point group symmetry on a surface of bulk material or of a single layer grown on a substrate is $C_{4v}$, and $s$- and $d_{x^2-y^2}$-wave belong to different representations of the point group ($A_1$ and $B_1$, respectively) and hence the spin-singlet pairing is either purely $s$-wave or purely $d_{x^2-y^2}$-wave.
However, in the orthorhombic phase ($a \neq b$ and $\gamma=90^\circ$, where $a$ and $b$ are the lengths of the lattice vectors and $\gamma$ is the angle between them) the four-fold rotation ($C_{4}$) and diagonal mirror reflection ($\sigma_d$ ) symmetries are broken, resulting in $C_{2v}$ as the reduced point group symmetry [see Tab.~\ref{tab:broken}].
In $C_{2v}$, $s$-wave and $d_{x^2-y^2}$-wave are no longer distinguished, while other representations ($A_2$, $B_2$, and $E$) of $C_{4v}$ stay symmetry-distinct from them.
The pairing gap in the nematic phase therefore is a mixture of $s$-wave and $d_{x^2-y^2}$-wave.
Within a Ginzburg-Landau (GL) theory, this mixing is described by
\begin{align}
  \calL_{s \cdot d} &= \eta s^* d_{x^2-y^2} + \mathrm{c.c.}
\end{align}
where $\eta$ is a real field representing the strength of nematicity, and $s$ and $d_{x^2-y^2}$ are the pairing order parameters in the corresponding channels.
Since $s$ is in the trivial $A_1$ representation of $C_{4v}$, while both $d_{x^2-y^2}$ and $\eta$ are in the $B_1$, their product is trivial.

\para{}%-------------------------------------------------------------
A domain wall between two nematic domains $a>b$ and $a<b$ forms a well-defined atomic junction when it is along the diagonal direction; it is indeed what is observed in FeSe~\cite{watashige_evidence_2015} and also in other iron-pnictides in the nematic phase~\cite{kalisky_stripes_2010}.
Under either $C_4$  or $\sigma_d$ symmetry operation $d_{x^2-y^2}$-wave component changes sign, while $s$-wave component is invariant.
Therefore, the relative sign between the mixed $s$- and $d_{x^2-y^2}$-wave components flips across the domain wall.
This change of relative sign between $s$-and $d_{x^2-y^2}$- component can manifest in two distinct ways: (1) the $s$-wave component stays constant or (2) the $d$-wave component stays constant across the domain wall. 
We refer to these two types of domain walls as $(d \pm s)$ and $(s \pm d)$, respectively.

\begin{figure}\centering
\subfigure[\label{fig:boundaryinduced_swave}]{
\includegraphics[]{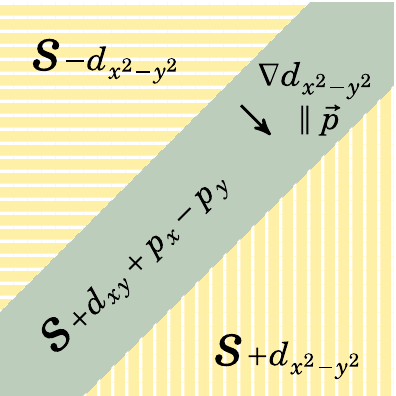}
}
\subfigure[\label{fig:boundaryinduced_dwave}]{
\includegraphics[]{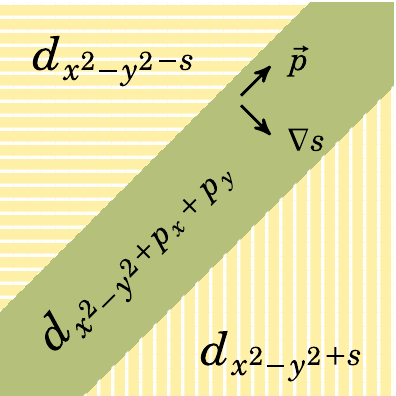}
}
\caption{\label{fig:boundaryinduced}%
Induced order parameter between two nematic domains.
\subref{fig:boundaryinduced_swave} $(s \pm d)$ domain wall across which $d_{x^2-y^2}$-wave component changes sign.
\subref{fig:boundaryinduced_dwave} $(d \pm s)$ domain wall across which $s$-wave component changes sign.
The induced $p$-wave component on the domain wall depends on which singlet order parameter changes sign.
}
\end{figure}

\para{}%-------------------------------------------------------------
A so far little noted fact is that additional pairing components can be mixed in locally at the domain wall because of the lower symmetry of the domain wall. On a diagonal domain wall parallel to the $[1\bar{1}0]$ plane as in Fig.~\ref{fig:boundaryinduced}, all the symmetry operations of $C_{4v}$ are broken except $\sigma_d(x-y)$: the mirror reflection with respect to the $[1\bar{1}0]$ plane [see Tab.~\ref{tab:broken}]. Since the component that remains finite at the domain wall behave distinctly under $\sigma_d(x-y)$ for the $(s\pm d)$ case [Fig.~\ref{fig:boundaryinduced_swave}, $s$ component is even] and the $(d\pm s)$ case [Fig.~\ref{fig:boundaryinduced_dwave}, $d_{x^2-y^2}$ component is odd], the new symmetry induced components also differ qualitatively for the two cases. Furthermore the SOC that is known to be substantial in FeSe~\cite{borisenko_direct_2016,ma_prominent_2016,cvetkovic_space_2013} has unusual implications at the domain wall.

\para{}%-------------------------------------------------------------
In the presence of the SOC, $\mathrm{SU(2)}$ spin-rotation symmetry is broken and singlets and triplets can mix in principle.
However, when the SOC conserves $S_z$, $C_2$ symmetry still present within each nematic domains forbids mixing between singlets and triplets in the bulk of the system. What has been overlooked so far is the fact that the domain wall itself lacks $C_2$ symmetry and hence a $S_z$ preserving SOC can couple $p$-wave components in the total $S_z=0$ channel with singlet components. This means the domain wall forms an emergent quasi-1D system with symmetry induced $p$-wave components.

\para{}%-------------------------------------------------------------
Now we consider the $(s\pm d)$ case and $(d\pm s)$ case separately.
For the $(s\pm d)$ case [see Fig.~\ref{fig:boundaryinduced_swave}], the component that stays finite at the domain wall is $s$-wave, which is even under $\sigma_d(x-y)$. Now the lack of the $C_2$ symmetry on the domain wall allows for  $d_{xy}$, which is also even under $\sigma_d(x-y)$, to mix in.
But most importantly a $p$-wave component perpendicular to the domain wall, $(-p_x+p_y)$-component, can be mixed in. One way to see this is to recognize that  $(-p_x+p_y)$ is also even under $\sigma_d(x-y)$ since $\sigma_d(x-y)$ affects both the spatial and spin coordinates of $p$-wave components. For instance, 
\begin{align}
  p_x
    &\sim
    c_{\hat{x}, \up} c_{-\hat{x}, \dn} +
    c_{\hat{x}, \dn} c_{-\hat{x}, \up},
    \label{eq:px}
\end{align}
where $\hat{x}$ is a spatial vector parallel to the $x$-axis transforms to $-p_y$ under  $\sigma_d(x-y)$. Similarly, $p_y$ transforms to $-p_x$. Hence $(-p_x+p_y)$ is even under $\sigma_d(x-y)$. Using the language of Ginzburg-Landau theory,
the $(s\pm d)$ domain wall imposes a gradient in the $d_{x^2-y^2}$ component perpendicular to the domain wall, i.e., $\nabla d_{x^2-y^2}\propto (1,-1)$. Since $p_y$ transforms as $-\hat{x}$ and $p_x$ transforms as $\hat{y}$, $(p_y^*\partial_x+p_x^*\partial_y)$ transforms as $B_1$ representation of $C_{4v}$. Hence the following coupling term is allowed by symmetry 
\begin{align}
  \calL_{s\pm d}
    &=      \gamma ( p_y^* \partial_x + p_x^* \partial_y ) d_{x^2-y^2}
      + \mathrm{c.c.}.
\end{align}
Now turning to the $(d\pm s)$ case [Fig.~\ref{fig:boundaryinduced_dwave}], the component that stays finite at the domain wall is now $d_{x^2-y^2}$, which is invariant under $\sigma_d(x-y)$ followed by a gauge transformation by $\pi$. From Eq.~\eqref{eq:px}, $p_x+p_y$ is also invariant under the same discrete transformation and hence now a $p$-wave component along the domain wall is induced [see Fig.~\ref{fig:boundaryinduced_dwave}].
From the GL theory perspective, it is the $s$-wave component that changes sign across the domain wall and hence the non-zero gradient is $\nabla s\propto (1,-1)$. $(p_y^*\partial_x-p_x^*\partial_y)$ transforms as $A_1$ representation, now the symmetry allowed coupling term is 
\begin{align}
  \calL_{d\pm s}
    &= \gamma ( p_y^* \partial_x  - p_x^* \partial_y ) s+ \mathrm{c.c.}
\end{align}
This implies that the gradient of $s$-wave component imposed by the domain wall in the dominantly $d_{x^2-y^2}$-wave nematic superconductor induces a $p$-wave component along the domain wall direction, defining an emergent 1D $p$-wave superconductor.

\begin{figure}\begin{center}%
\subfigure[\label{fig:induced_sp}]{\includegraphics[height=1.55in]{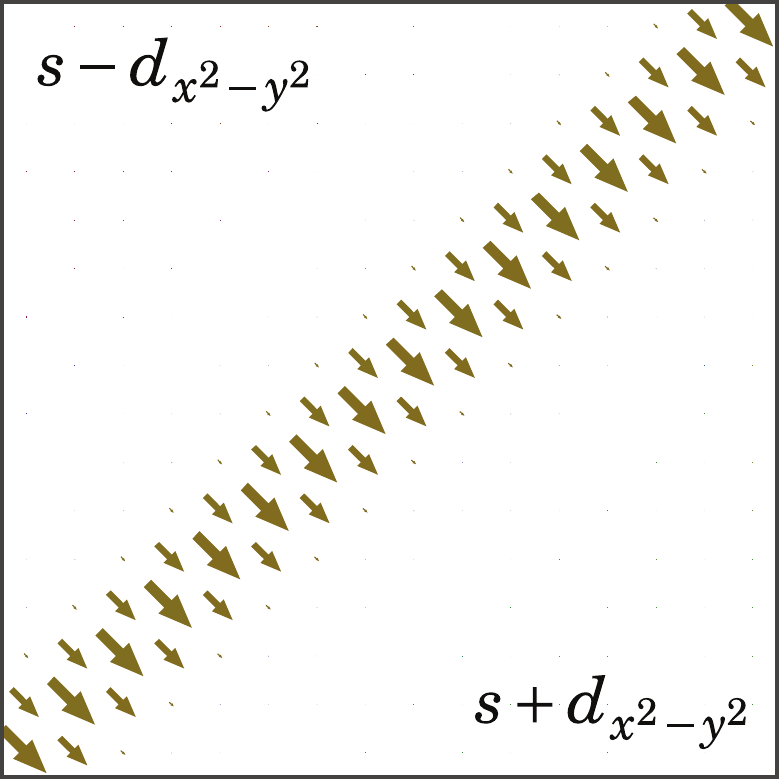}}
\subfigure[\label{fig:induced_dp}]{\includegraphics[height=1.55in]{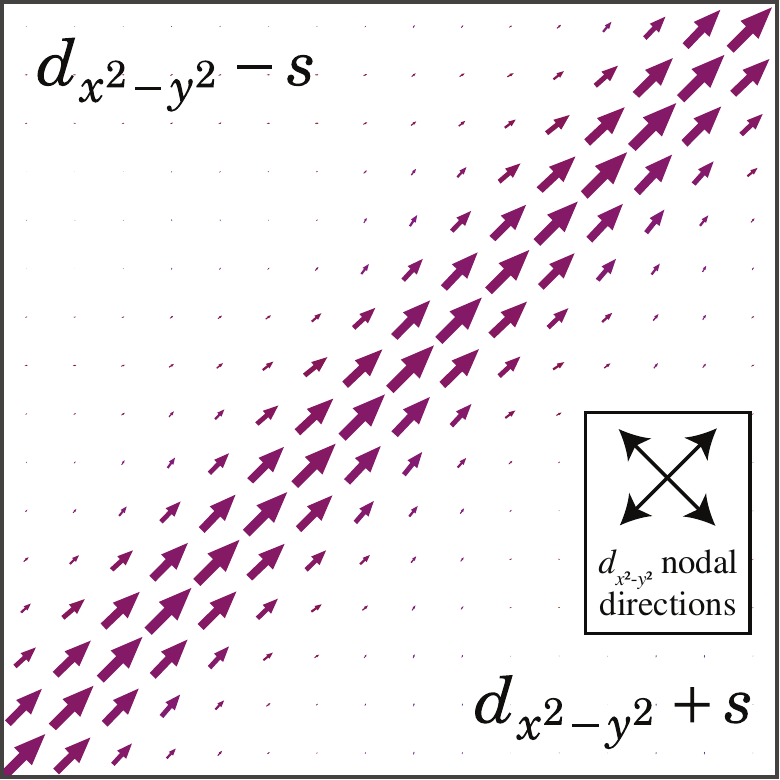}}
\caption{\label{fig:induced}%
Spatial distribution of pair amplitude in $p$-wave channel measured across a domain wall \subref{fig:induced_sp} when uniform $s$-wave pairing term is imposed, and \subref{fig:induced_dp} when uniform $d_{x^2-y^2}$-wave pairing is imposed.
The direction of each arrow in \subref{fig:induced_sp} and \subref{fig:induced_dp} represent the direction of $\bfp(\bfr)$ component at each point as defined in Eq.~\eqref{eq:component_p}.
The inset of \subref{fig:induced_dp} shows the nodal directions of the $d_{x^2-y^2}$-wave component.
Here we set
$\mu = 0.2$, $\lambda=1$, $t_{xz, xz}(\hat{x}) = 1$, $t_{xz, xz}(\hat{y}) = -0.5$, $t_{xz, xz}(\hat{x}+\hat{y}) = t_{xz, yz} (\hat{x}+\hat{y}) = 0.2$, $\Delta=0.2$, and $\eta_0=0.2$.
}
\end{center}\end{figure}

\para{}%-------------------------------------------------------------
The above symmetry-based insights can be readily confirmed through an explicit  microscopic calculation. 
For demonstration purpose, consider a two band toy model whose hopping is given in terms of operators $c_{\alpha, \sigma}(\bfr)$ which annihilate an electron at site $\bfr$ with orbital $\alpha=xz, yz$ and spin $\sigma=\up, \dn$ as
\begin{align}
\label{eq:kinetic}
  H_{\mathrm{kinetic}}(\bfr, \bfr')
    &=
      \bft(\bfr-\bfr') \sigma_0
      -\mu\delta_{\bfr,\bfr'} \tau_0 \sigma_0 
-\lambda \delta_{\bfr, \bfr'} \tau_2 \sigma_3
\end{align}
where $\tau_i$ and $\sigma_i$ for $i=0,1,2,3$ are identity and Pauli matrices operating on orbital and spin spaces, respectively.
Here, $\mu$ and $\lambda$ chemical potential, and spin-orbit coupling, and  $\bft(\boldsymbol{\rho})$ is a matrix in orbital space which parametrizes hopping.
In addition, we impose uniform singlet pairing of the dominant component for each domain wall configuration through
\begin{align}
  \calH_{\mathrm{pair}}
    &=
      \Delta
      \sum_{\bfr}
      \sum_{\alpha=xz, yz }
          f_{\alpha}
          c_{\alpha,\up}^\dag (\bfr) c_{\alpha,\dn}^\dag (\bfr)
        + \mathrm{H.c.}
\label{eq:pair}
\end{align}
where $f_{\alpha}$ is the orbital form factor: $f_{\alpha} = f^{s}_{\alpha} \equiv 1$ for $s$-wave, and $f_{\alpha} = f^{d_{x^2-y^2}}_{\alpha} \equiv [\tau_3]_{\alpha,\alpha}$ for $d_{x^2-y^2}$-wave. For $(s\pm d)$ domain wall we impose a uniform $s$-wave pairing and for $(d\pm s)$ domain wall we impose a uniform $d_{x^2-y^2}$ pairing.

\para{} 
Now we impose a sharp nematic domain wall profile through an on-site orbital imbalance that changes sign across the domain wall, i.e.,
\begin{align}
 \label{eq:nematic}
  H_{\mathrm{nematic}}(\bfr, \bfr')
    &=
      \delta_{\bfr, \bfr'} \eta(\bfr) \tau_3 \sigma_0.
\end{align}
where $\eta(x,y) = \eta_0 (2 \Theta(x-y) - 1)$.
We then solve this mean-field theory to obtain the Bogoliubov eigenstates and measure the pair amplitudes on sites and nearest neighbor bonds with the obtained eigenstates. Without nematicity defined in Eq.~\eqref{eq:nematic}, the measured pair amplitudes will trivially follow the symmetry of the imposed uniform pairing in Eq.~\eqref{eq:pair}. But the imposed nematic domain wall induces secondary components both in the domains as well as on the domain wall. In particular, plots of $S_z=0$ spin-triplet components defined by 
\begin{align}
\label{eq:component_p}
  \bfp(\bfr)
    &=
      \sum_{\substack{\alpha=xz,yz \\ \boldsymbol{\rho} = \pm \hat{x}, \pm \hat{y}}}
        \boldsymbol{\rho}
        \langle
          c_{\alpha,\up}(\bfr + \boldsymbol{\rho}) c_{\alpha,\dn}(\bfr) +
          (\up\; \longleftrightarrow \;\dn)
        \rangle.
\end{align}
in Fig.~\ref{fig:induced_sp} clearly shows $p$-wave components concentrated on the domain walls with its direction perpendicular to the domain wall for $(s\pm d)$ case and parallel to the domain wall for $(d\pm s)$ case.

\begin{figure}
\centering
\subfigure[\label{fig:ldosexp}]{\includegraphics[height=1.62in]{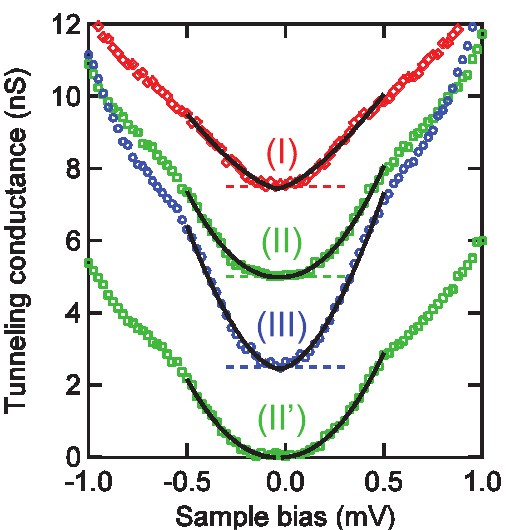}}
\subfigure[\label{fig:ldos2}]{\includegraphics[height=1.62in]{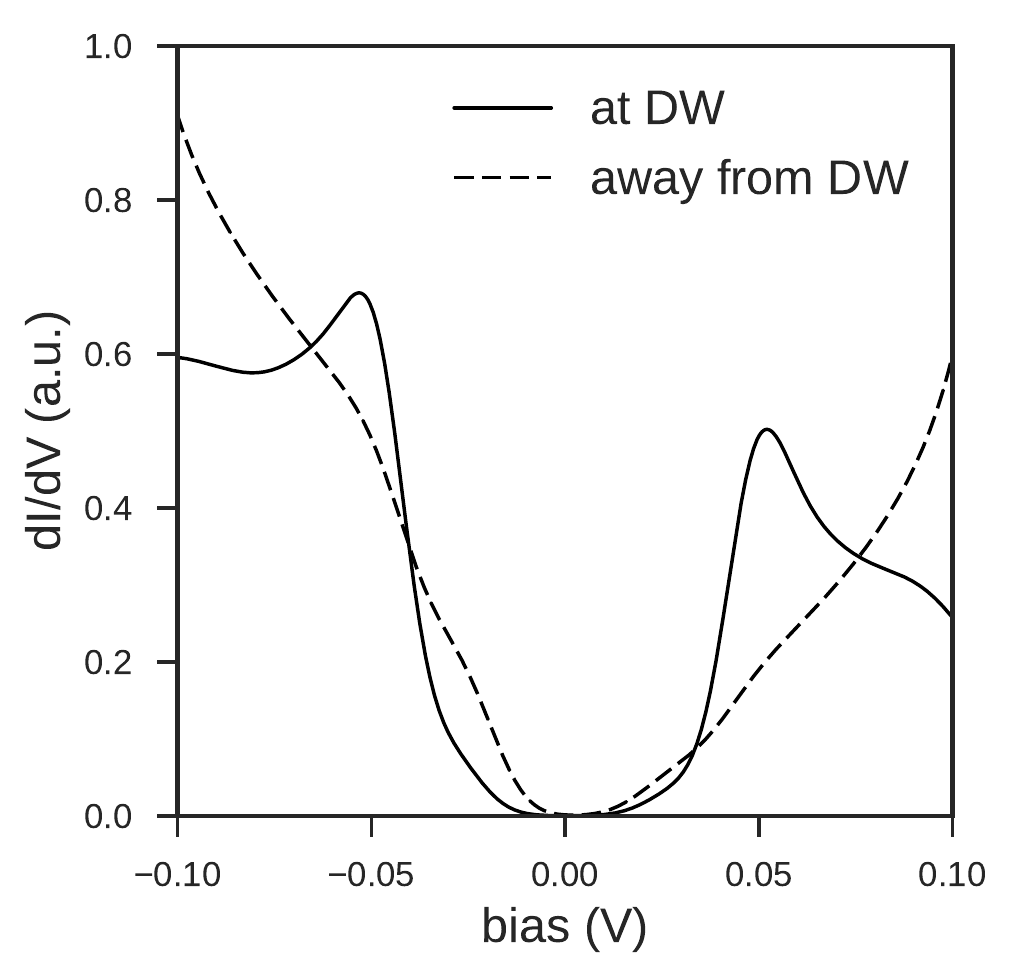}}
\caption{\label{fig:ldos}
    \subref{fig:ldosexp}
    Local tunneling conductance at/away from the domain wall
    measured by scanning tunneling spectroscopy (reproduced from Ref.~\cite{watashige_evidence_2015}).
    (I) indicates local tunneling conductance measured away from a domain wall, (II) and (II') near a domain wall, and (III) at a domain wall.
    \subref{fig:ldos2} local tunneling conductance calculated with induced sin-triplet pairing ($\Delta_{p} = 2$) at the domain wall.
    The three-orbital model by \textcite{daghofer_three_2010} has been used as the underlying band structure. The magnitudes of $d$-wave, which is uniform, and the extended $s$-wave, which changes sign across the domain wall, are chosen to be $\Delta_{d}=0.05$, $\Delta_{s}=0.025$, respectively.
}
\end{figure}

\para{}
A scanning tunneling spectroscopy experiment found suppression of low energy density of states near nematic domain walls~\cite{watashige_evidence_2015}, as shown in Fig.~\ref{fig:ldosexp}.
The authors suggested that the suppresion indicates enhancement of superconducting gap, which they conjecture is due to spin-triplet pairing induced by spontaneous time-reversal breaking.
Here we point out that spin-triplet pairing induced by a combination of spatial gradient of singlet pairing order parameter and spin-orbit coupling provides an alternative explanation for the suppresion of low energy states.
For concreteness, we computed the tunneling conductance at the domain wall and away from domain wall, with and without the additional spin-triplet component.
As shown in Fig.~\ref{fig:ldos} in the presence of induced spin-triplet pairing at the domain wall, the low energy density of states is suppressed.
As a result, the tunneling conductance appears more `U'-shaped compared to that of the bulk.
Such `U'-shape tunneling conductance is consistent with what is measured in Ref.~\cite{watashige_evidence_2015}.

\para{}
Interestingly, the $(s\pm d)$ case can be viewed as two edges of $d$-wave superconductors brought close to each other. From this view point, the induced $p$-wave perpendicular to the domain wall is a way a pair of flat-band zero modes predicted in Refs.~
\cite{potter_edge_2014,hofmann_edge_2016} pair up to gap out the low energy spectrum.  Alternatively a large $p$-wave component parallel to the nematic domain wall can defining an emergent one-dimensional topological superconductor for the $(d\pm s)$ case.
From the symmetry classification perspective~\cite{schnyder_classification_2008}, our system belongs to the DIII class with time reversal symmetry as in Ref.~\cite{PhysRevB.88.134523} with one difference being that our block Hamiltonians for each $S_z$ blocks belong to AIII class. Hence the $(d\pm s)$ opens possibility for zero energy Majorana bound states at the end of the ``wire''.

\begin{figure}
\subfigure[\label{fig:boundstate2d-without-p}]{%
\includegraphics[height=0.95in]{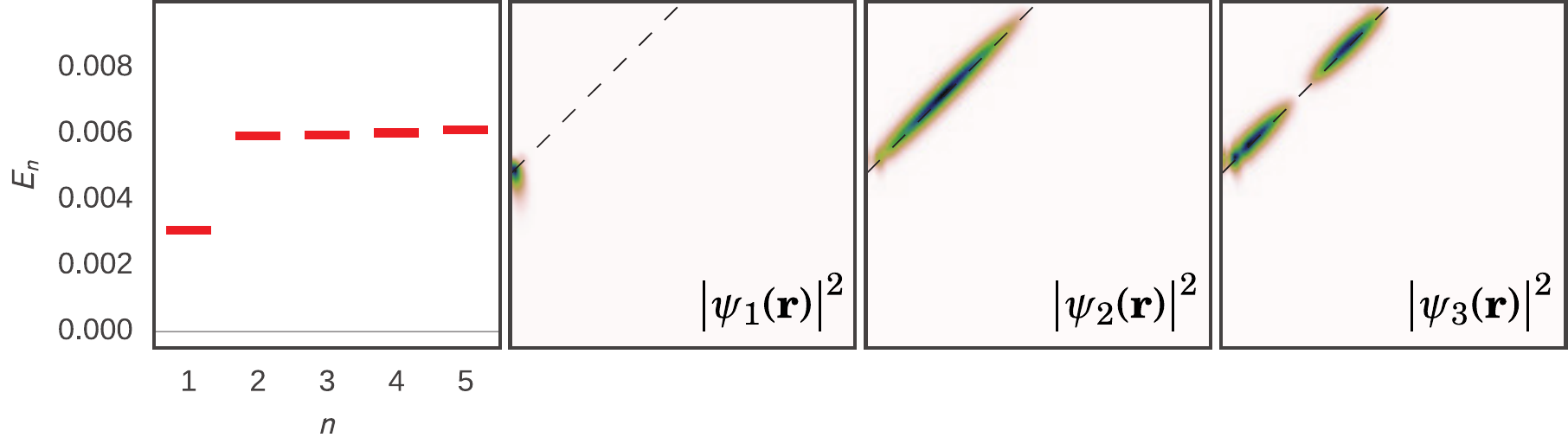}
}\\
\subfigure[\label{fig:boundstate2d-with-p}]{%
\includegraphics[height=0.95in]{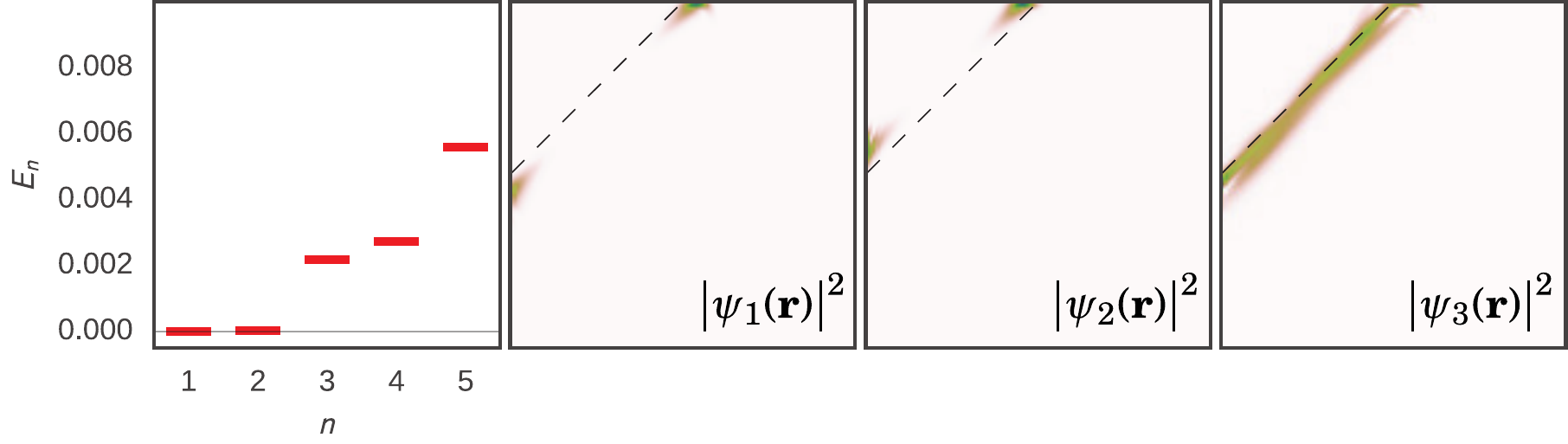}%
}
\caption{\label{fig:boundstate2d}%
Low lying eigenstates of a BdG Hamiltonian with a $(d\pm s)$ domain wall for $n$'th positive energy eigenstates \subref{fig:boundstate2d-without-p} without $p$-wave component, and \subref{fig:boundstate2d-with-p} with $p$-wave component at the domain wall.
The spectra (the left most panels) and the spatial distribution of wave-function amplitudes $|\psi_n(\bfr)|^2$ of $n=1,2$ and $3$.
The dashed lines indicate the location of the domain wall.
}
\end{figure}

\para{}%-------------------------------------------------------------
To demonstrate the implication of the emergent $p$-wave wire on the $(d\pm s)$ domain wall, we now consider a simple one-band Bogoliubov-de Gennes Hamiltonian.
We consider two limits: when $p$-wave component is zero and when it is large.
We work with a Bogoliubov-de Gennes Hamiltonian
\begin{align}
  H_{\bfr,\bfr'}
    &=
      \begin{pmatrix}
        t_{\bfr, \bfr'}          &  \Delta_{\bfr, \bfr'} \\
        \Delta_{\bfr', \bfr}^{*} & -t_{\bfr, \bfr'}
      \end{pmatrix}.
\label{eq:boundstate2d-hamiltonian}
\end{align}
The hopping includes chemical potential and nearest neighbor hoppings $t_{\bfr, \bfr'} = -\mu \delta_{\bfr, \bfr'} - t \delta_{\langle \bfr, \bfr' \rangle}$, with $\mu = -3$ and $t=1$.
The imposed pairing is a superposition of uniform $d_{x^2-y^2}$-wave with domain-defining $s$-wave component (changing sign across the domain wall) and induced $p$-wave component local to the domain wall.
The low energy spectra of the domain wall with and without the induced $p$-wave pairing differ qualitatively.
Figures~\ref{fig:boundstate2d-without-p} and \ref{fig:boundstate2d-with-p} show the energies of the low-lying single particle excitations near a $(d\pm s)$ domain wall and their spatial profile, without and with the induced $p$-wave components, respectively.
In both cases, we have set $\Delta_d = 0.6$, $\Delta_s = 0.4$, and $\xi=8$.
The $p$-wave component $\Delta_p$ is set to zero and 150 in Figs.~\ref{fig:boundstate2d-without-p} and \ref{fig:boundstate2d-with-p}, respectively.
In the leftmost panels of Figs.~\ref{fig:boundstate2d-without-p} and \ref{fig:boundstate2d-with-p}, we plot positive the eigenenergies, since the product of time-reversal and particle-hole symmetry ensures that the eigenenergies come in $(E, -E)$ pairs.
Without the $p$-wave component on the domain wall, the excitation gap remains non-zero.
When large $p$-wave component is introduced, on the other hand, energies of the first two eigen states drop to a value indistinguishable from zero within our calculation.
These zero modes peak at $\sim \xi$ away from the center of the domain wall on both of its sides, which account for the multiplicity of two for each spin state. Hence our domain wall supports four-channels of spinless $p$-wave wires.

\para{}%-------------------------------------------------------------
Interestingly, the symmetry analysis we used here applies to a related situation of a clean edge of a $d$-wave superconductor~\cite{potter_edge_2014,hofmann_edge_2016}. It was shown in Ref.~\cite{hofmann_edge_2016} using quantum Monte Carlo simulations that ferromagnetic instability of Majorana  flat band at a [110] edge of a $d_{x^2-y^2}$ discussed by \textcite{potter_edge_2014} accompanies $p$-wave pairing along the edge. From the GL-theory perspective we have been using throughout this letter,
\begin{align}
\calL_{M-t} &\propto
    M_z \left( p_x^* \partial_x - p_y^* \partial_y \right) d_{x^2-y^2}
    + \mathrm{c.c.}
\end{align}
is a symmetry allowed term. This is because the magnetization order parameter $M_z$ belongs to $A_2$ representation while $p_x^* \partial_x - p_y^* \partial_y$, and $d_{x^2-y^2}$ 
respectively falls into $B_2$, and $B_1$ representations.
Since the product is trivial, such a term is allowed and non-zero magnetization will be accompanied by $p$-wave pairing along the edge. This precedence of correspondence between exact numerical solution and our symmetry analysis lends further confidence to our predictions.

\para{}%-------------------------------------------------------------
To summarize, we considered the problem of singlet-triplet mixing on a nematic domain wall in a superconductor with $S_z$ preserving spin-orbit coupling with FeSe in mind. 
First we noted that $C_2$ symmetry of each nematic domain requires the order parameter representation to be a mix of $d$-wave and $s$-wave component. Further lowering of symmetry on the domain wall allows for spin triplet components to mix in. 
We then noted that two distinct realizations of domain boundaries are possible depending on the dominant order parameter component. Specifically when the $d$-wave component is more dominant, the $s$-wave component changes sign across the domain wall($d\pm s$ domain wall) whereas when the $s$-wave component is more dominant the $d$-wave component changes sign across the domain wall ($s\pm d$ domain wall). The two types of domain wall each support a locally induced spin-triplet $p$-wave components along different directions: $p$-wave along the wall direction for the $(d\pm s)$ domain wall and $p$-wave perpendicular to the wall direction for the $(s\pm d)$ domain wall. 
The $p$-wave component aligned parallel to the domain wall raises a tantalizing possibility of realizing a emergent 1D triplet superconductor with only four channels. Indeed our Bogoliubov-de Gennes calculation shows that such emergent 1D triplet superconductor will support Majorana zero-energy bound states. Given growing understanding of significance of $d$-component in FeSe~\cite{huang,Hirschfeld15PRL,Sprau16}, our findings call for further investigation of domain walls in FeSe and a search for Majorana zero modes at higher temperatures.

\vspace{5mm}
\noindent {\bf Acknowledgements} We thank J.C. Davis, Jason Alicea, Rafael Fernandes, Piers Coleman, and  Jian-Huang She for discussions. E-AK  and KL were  supported by the U.S. Department of Energy, Office of Basic Energy Sciences,  Division of Materials Science and Engineering  under  Award  DE-SC0010313.   E-AK  also  acknowledges Simons Fellow in Theoretical Physics Award\#392182  hospitality of the KITP supported by Grant No. NSF PHY11-25915. KL was supported in part by DOE Grant DE-FG02-07ER46423.

\end{document}